\documentclass[12pt]{article}
\usepackage{sao1}
\usepackage{graphics,amsmath,amssymb}

\newcommand{\de}{{\mathrm d}}

\newcommand{\zav}[1]{\left(#1\right)}
\newcommand{\hzav}[1]{\left[#1\right]}

\newcommand{\vr}{{v_r}}

\newcommand{\NR}{\mathrm{N \! R}}

\begin{document}

\title{Multicomponent stellar winds of He chemically peculiar stars}

\author{Krti\v cka J. \inst{1,2} \and Kub\'at J. \inst{2}}

\institute{
\'Ustav teoretick\'e fyziky a astrofyziky P\v{r}F~MU,
Kotl\'a\v{r}sk\'a 2, CZ-611~37 Brno, Czech~Republic
\and Astronomick\'y \'ustav, AV \v{C}R, CZ-251 65 Ond\v{r}ejov,
Czech~Republic
}

\maketitle 

\begin{abstract}
We calculate multicomponent stellar wind models with inclusion of a
helium component applicable to He-rich and He-poor stars.
We show that helium does not decouple from the stellar wind of He-rich
stars due to its coupling to hydrogen and its ionized state.
For He-poor stars helium may decouple from the stellar wind, however
this effect is not able to explain the chemical peculiarity of these stars.
We conclude that the explanation of chemical peculiarity of these stars
based purely on helium or hydrogen decoupling from the stellar wind is unlikely.
\end{abstract}


\section{Introduction}

Radiative force may have important influence on the structure of
stellar atmospheres. Although there is not known any significant
direct influence of the radiative force on the atmospheres of most cool
stars, the radiative force may be dominant in the atmospheres of hot stars.
The consequences of this influence are different. Mainly for A stars
and white dwarfs the radiative force may cause the radiative diffusion and
subsequent
elemental abundance anomalies (see Alecian 1995 or Vauclair 2003 for a
review). On the other hand, the radiative force may accelerate
a stellar wind mainly for O stars and WR stars (Kudritzki \& Puls 2000).
However, there exists a group of B stars, for which both effects of
stellar wind and elemental diffusion in the stellar atmosphere are
important. It is generally believed that in order to explain the elemental
abundances of helium and some other elements in the atmospheres of
He-rich stars at least three different ingredients are necessary.
These ingredients are stellar wind, elemental diffusion and magnetic
field (e.g. Michaud et al. 1987).

However, alternative explanation of chemical peculiarity of He-rich and He-week
stars exists. It was proposed by Hunger \& Groote (1999, hereafter HG).
They showed that for stars with effective temperatures
$T_\mathrm{eff}<25\,000\,\mathrm{K}$ helium may decouple from the
stellar wind, fall back onto the stellar surface and create regions with
enhanced abundance of helium if magnetic fields are present.
Similarly, for stars with effective temperatures
$T_\mathrm{eff}<17\,000\,\mathrm{K}$ even hydrogen may decouple from the
stellar wind, fall back onto the stellar surface and cause helium
underabundance if magnetic fields are present.
We decided to test this explanation of chemical peculiarity of He-rich
and He-week stars using our multicomponent wind models.
First preliminary calculations have already been performed
by Krti\v{c}ka \& Kub\'at (2001a), however with artificially low helium
charge and without ionization balance calculation.

\section{Multicomponent stellar wind models}

Stellar winds of hot stars are accelerated mainly by the absorption of radiation
in the resonance lines of such elements like carbon, nitrogen, oxygen, or iron.
However, these wind components have much lower density than the
rest of the stellar wind, which is composed mainly of hydrogen and helium. Thus,
the process of acceleration of a stellar wind of hot stars has two different
physical steps.
In the first step, radiation is absorbed by low-density metals like
carbon, nitrogen, oxygen, or it
is scattered by free electrons, and these wind components obtain
momentum from stellar radiation field.
In the second step, momentum obtained by these low-density components is
transferred to high-density wind components via the Coulomb collisions
between charged particles. Thus, stellar winds
of hot stars have a multicomponent nature.
For stars with relatively high wind density (e.g. galactic O stars or
WR stars) this multicomponent wind nature does not influence wind
structure, as has been discussed by Castor et al. (1976).
However, for stars with a relatively low wind density (i.e.
main-sequence B stars or stars with extremely low metallicity, see
Krti\v{c}ka et al. 2003) the multicomponent wind nature becomes
important for the wind structure itself because new effects occur,
for example frictional heating or decoupling of wind components
(Springmann \& Pauldrach 1992, Babel 1995, Krti\v{c}ka \& Kub\' at
2001b, hereafter KKII).

\subsection{Model equations}

Equations used here
for the calculation of four-component wind models are nearly the
same as those of KKII.
We assume stationary and spherically symmetric stellar wind, which is
composed of four components, namely
absorbing metallic ions, hydrogen, helium, and free electrons.
For the calculation of wind models we solve the
continuity equation, momentum equation, and energy equation for each
component of the flow. The continuity equation has the form
\begin{equation}
\label{kontrov}
\frac{\de }{\de  r}\zav{r^2\rho_a{\vr}_a}  =  0,
\end{equation}
where $\rho_a$ is the density, $r$ is the radius, and $\vr_a$
is the velocity of a component $a$. The momentum equation is
\begin{equation}
\label{pohrov}
{\vr}_a\frac{\de {\vr}_a}{\de r}=
{g}_{a}^{\mathrm{rad}}-g-\frac{1}{{\rho}_a}\frac{\de }{\de
r}\zav{{a}_a^2{\rho}_a}
 +\frac{q_a}{m_a}E +
 \sum_{b\neq a} g^{\mathrm{fric}}_{ab},
\end{equation}
where ${g}_{a}^{\mathrm{rad}}$ is the radiative force,
$E$ is electric polarisation field, and
$g^{\mathrm{fric}}_{ab}$ is the frictional force (Burgers 1969)
\begin{equation}
\label{treni}
g^{\mathrm{fric}}_{ab}=
\frac{1}{{\rho}_a}K_{ab}G(x_{ab})\frac{{\vr}_b-{\vr}_a}{|{\vr}_b-{\vr}_a|},
\end{equation}
where $G(x_{ab})$ is the Chandrasekhar function, and the frictional coefficient is
\begin{equation}
\label{trecko}
{K}_{ab}={n}_a{n}_b\frac{4\pi  {q}_a^2{q}_b^2}{k T_{ab}}\ln\Lambda,
\end{equation}
where mean temperature $T_{ab}=\zav{m_b T_a + m_a T_b}/\zav{m_b+m_a}$ is
calculated using temperatures $T_a$ and $T_b$ of individual wind
components with atomic masses $m_a$ and $m_b$.
The radiative force in the CAK approximation (Castor et al. 1975,
Friend \& Abbott 1986, Pauldrach et al. 1986,
see also KKII for generalization of CAK force for a multicomponent flow)
due to line-absorption acts on metals and radiative force due to the
Thomson scattering acts on free electrons.

Energy equation for each component of the flow is
\begin{equation}
\label{teprov}
\frac{3}{2}\vr_a{\rho_a}\frac{\de a^2_a}{\de r}+
\frac{a_a^2\rho_a}{r^2}\frac{\de }{\de r}\zav{r^2 {\vr}_a}=
Q_a^{\mathrm{rad}}+\!\sum_{b\neq a}\!\!\zav{ Q^{\mathrm{ex}}_{ab}\!+\!
Q^{\mathrm{fric}}_{ab}},
\end{equation}
where the heat exchange is given by
\begin{equation}
Q^{\mathrm{ex}}_{ab}=
\frac{1}{\sqrt\pi}K_{ab} \frac{2k\zav{T_b-T_a}}{m_a+m_b}
 \frac{\exp\zav{-x_{ab}^2}}{\alpha_{ab}},
\end{equation}
frictional heating is
\begin{equation}
Q^{\mathrm{fric}}_{ab}=
 \frac{m_b}{m_a+m_b}K_{ab}G(x_{ab})|{\vr}_b-{\vr}_a|,
\end{equation}
and $Q_a^{\mathrm{rad}}$ is the radiative heating
calculated using the
the method of thermal balance of electrons
(Kub\'at et al. 1999).
We also
take into account the
Gayley-Owocki (1994) heating.
The base flux,
which is
necessary for the calculation of the radiative heating term,
is taken from H-He spherically symmetric NLTE model atmospheres
(Kub\'at 2003).

System of hydrodynamic equations is closed using equation for electric
polarisation field
and by equations of ionization equilibrium
(we assume nebular
approximation after Mihalas,
1978, Eq. 5.46).
For a more detailed description of three-component variant of these
models see KKII.

\subsection{The possibility of wind decoupling}

\begin{figure}
\centering
\resizebox{0.7\hsize}{!}{\includegraphics{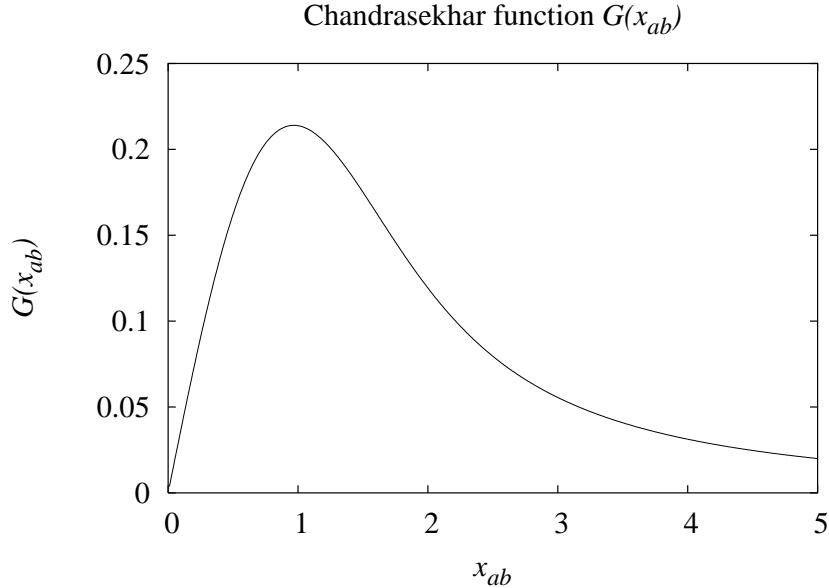}}
\caption[]{
{\bf \em The run of Chandrasekhar function.}
Note
that Eq.~(\ref{defxpi}) yields $x_{ab}\sim\Delta v_{ab}$, where
$\Delta v_{ab}$ is the velocity difference between wind
components.
If
the flow is well coupled, $x_{ab}\lesssim 1.0$,
$G(\Delta v_{ab})\sim \Delta v_{ab}$.
If
the drift velocity is large, $x_{ab}\gtrsim 1.0$,
$G(\Delta v_{ab})\sim \Delta v_{ab}^{-2}$, and wind may decouple.
Note
that the point $x_{ab} \approx 1.0$ corresponds to the
maximum of $G$.}
\label{chandra}
\end{figure}

The frictional force Eq.\,(\ref{treni}) between two components depends
on the velocity difference between these two components via the
so-called Chandrasekhar function,
\begin{equation}
G(x_{ab})=\frac{1}{2
x_{ab}^2}\hzav{\Phi(x_{ab})-x_{ab}\frac{\de\Phi(x_{ab})}{\de x_{ab}}},
\end{equation}
where non-dimensional velocity difference $x_{ab}$ is given by
\begin{equation}
\label{defxpi}
x_{ab}=\frac{|{\vr}_b-{\vr}_a|}{\alpha_{ab}}
=\frac{\Delta v_{ab}}{\alpha_{ab}},
\end{equation}
where $\alpha_{ab}$ is the mean thermal speed,
$\alpha_{ab}^2={2k\zav{m_aT_b+m_bT_a}}/\zav {m_am_b}$.
The plot of
the
Chandrasekhar function is given in Fig.\ref{chandra}.
For relatively low velocity
differences, 
$\Delta v_{ab}\lesssim \alpha_{ab}$,
the Chandrasekhar
function is
an
increasing function of the velocity difference 
$\Delta v_{ab}$
and components $a$ and $b$ are well coupled.
However, for higher velocity differences,
$\Delta v_{ab}\gtrsim \alpha_{ab}$,
the Chandrasekhar function is a
decreasing function of the velocity difference and the decoupling of
components may occur.

\subsection{Solution using Newton--Raphson method}

Multicomponent wind
equations
(\ref{kontrov},\ref{pohrov},\ref{teprov})
can be formally written as
\begin{equation}
\mathsf{P}\boldsymbol\psi = 0,
\end{equation}
where the vector describing the solution has the form of
\begin{equation}
\boldsymbol\psi=\zav{\boldsymbol\psi_1,\boldsymbol\psi_2,\dots,\boldsymbol\psi_\NR}^{\mathrm{T}},
\end{equation}
where
\begin{equation}
\boldsymbol\psi_i=\zav{\zav{\rho_{a,i}, v_{ra,i},
T_{a,i},z_{a,i}}_{a=\text{i,\,p,\,e}},
E_i,\Delta v_{r,i}}.
\end{equation}
The velocity difference $\Delta v_{r,i}$ was taken as an additional
independent variable to improve convergence properties.
The solution can be obtained using
the iteration scheme
\begin{equation}
\mathsf{J}^{n}\delta\boldsymbol\psi^{n+1}=-\mathsf{P}^{n}\boldsymbol\psi^{n},
\end{equation}
where the Jacobi-matrix is
\begin{equation}
{J}_{kl}^{n}=\frac{\partial P_{k}}{\partial\psi_l}.
\end{equation}
More
details
about the solution procedure can be found in Krti\v{c}ka (2003).

\section{Calculated models}

We have calculated four-component stellar wind models (i.e. models
consisting
of
absorbing ions, hydrogen, helium, and free electrons) to test the
explanation of helium chemically peculiar stars proposed by HG.
Here we present two stellar wind models of B3 and B6 main-sequence
stars. 
Spectral type of helium-rich stars is close to B3, for which HG
predicted helium decoupling from the stellar wind.
On the other hand, spectral type of helium-poor stars is close to B6.
For these stars HG predicted hydrogen decoupling from the stellar wind.

\subsection{Models suitable for He-rich stars}

\begin{figure}[t]
\centering
\resizebox{0.49\hsize}{!}{\includegraphics{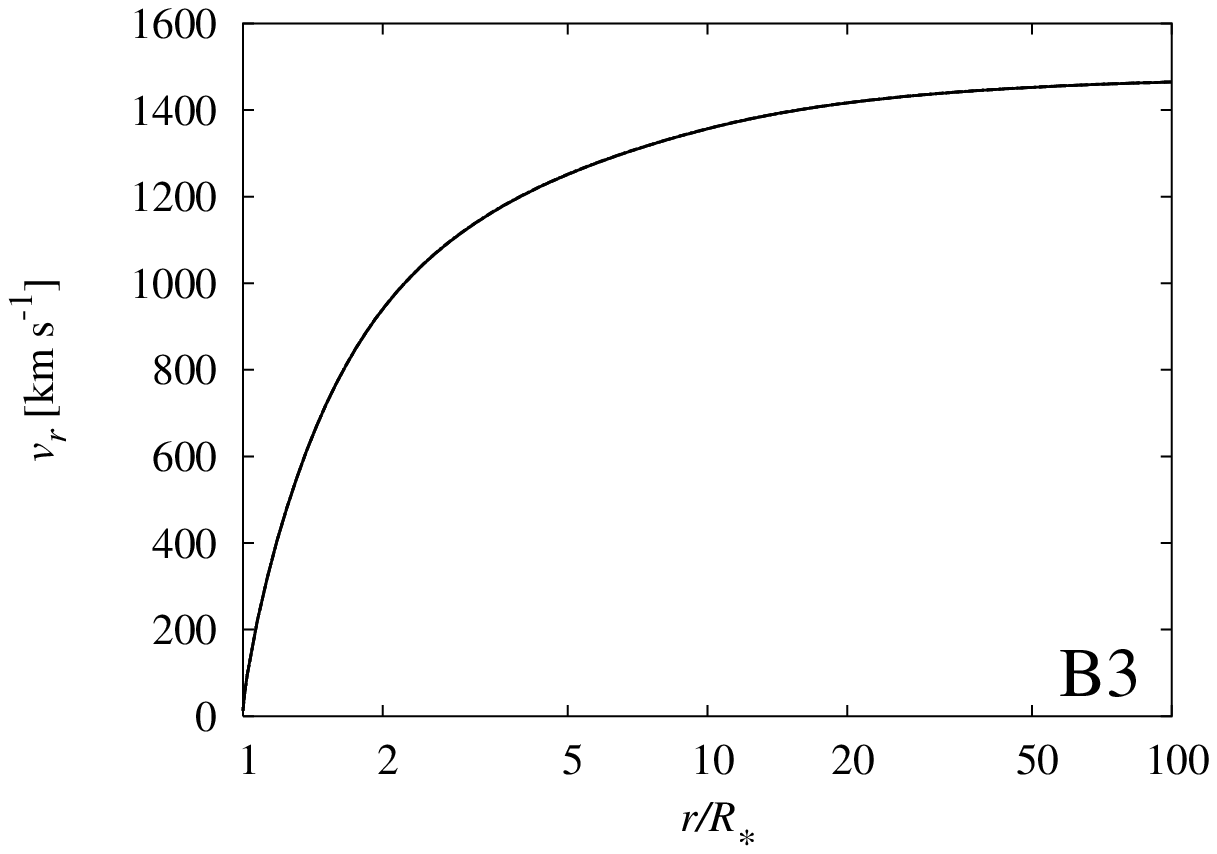}}
\resizebox{0.49\hsize}{!}{\includegraphics{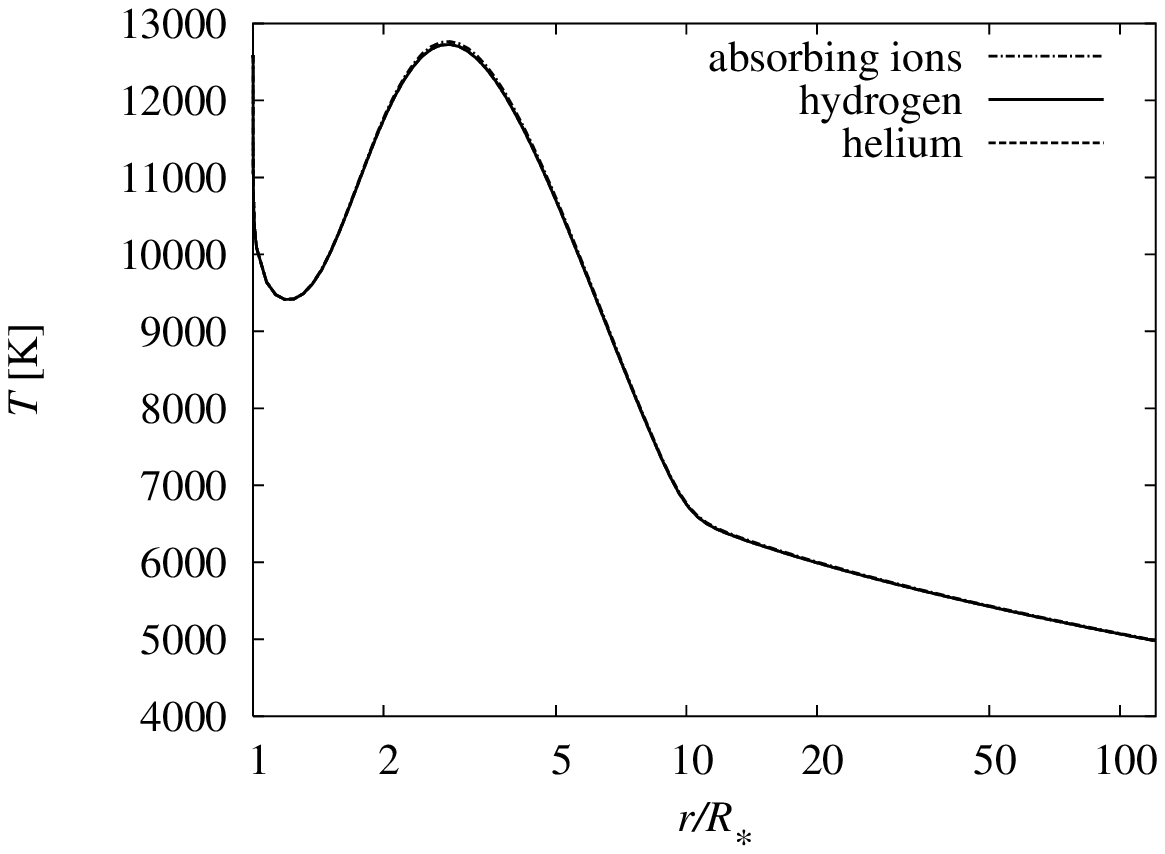}}
\resizebox{0.49\hsize}{!}{\includegraphics{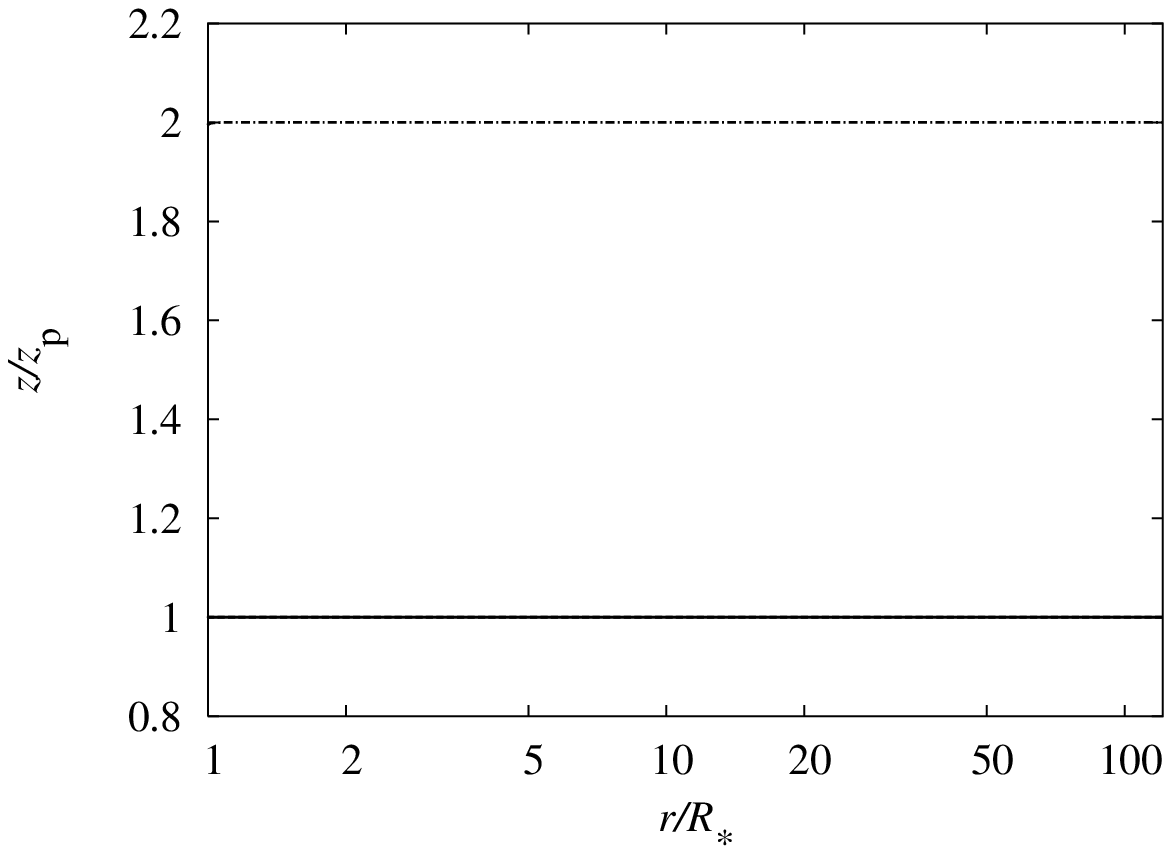}}
\caption[]{
{\bf \em Four-component wind model of a main-sequence B3 star.}
{\em Upper left panel:} Wind velocities of wind components.
Note
that velocities of all wind components are nearly the same.
{\em Upper right panel:} Temperatures of wind components.
Temperatures of all wind components are close to each other.
Note slight heating due to the Gayley \& Owocki (1994) and frictional
heating.
{\em Lower panel:} Charges of wind components.}
\label{b3}
\end{figure}

We have selected a main-sequence B3 star as a representative
example of
He-rich stars.
Stellar parameters are taken
according to
Harmanec (1988) and force multipliers are after Abbott (1982).
The plot of velocities, temperatures and charges of
individual wind components is given in Fig.\,\ref{b3}.

Contrary to the predictions of HG, helium does not decouple from the
main wind.
The reason for this behaviour is a neglect of helium coupling to
hydrogen by HG.
They
assumed that helium is accelerated mainly due to the collisions with
metallic (absorbing)
ions. The
frictional force 
$f_{\mathrm{He,i}}=\rho_{\mathrm{He}}g_{\mathrm{He,i}}$ is proportional
to (see Eqs.\,(\ref{treni},\ref{trecko}))
$$f_{\mathrm{He,i}}\sim n_\mathrm{He} n_\mathrm{i},$$
where $n_\mathrm{He}$ and $n_\mathrm{i}$ are number densities of helium
and metallic ions.
However, the frictional force due to the collisions between helium and
hydrogen atoms is
$$f_{\mathrm{He,H}}\sim n_\mathrm{He} n_\mathrm{H},$$
where $n_\mathrm{H}$ is the number density of hydrogen atoms.
Since
$n_\mathrm{H}\gg n_\mathrm{i}$,
the frictional force due to hydrogen
atoms may be important for the acceleration of helium atoms, and,
consequently,
can not be neglected.

We conclude that helium does not decouple from the stellar wind of
He-rich stars.
Thus, the explanation of
a helium overabundance in
chemically peculiar stars by the helium
decoupling in their stellar wind is 
questionable.

\subsection{Models suitable for He-poor stars} 

\begin{figure}[t] 
\centering
\resizebox{0.49\hsize}{!}{\includegraphics{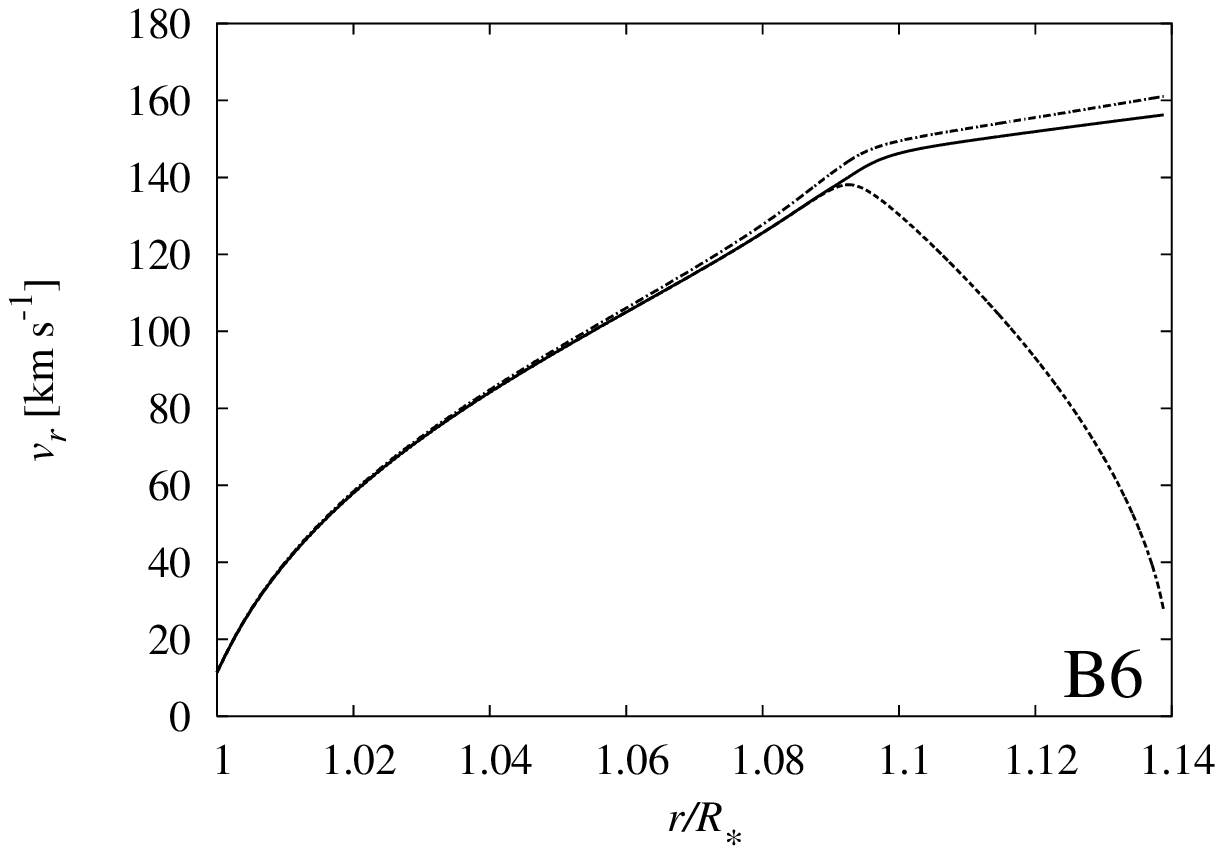}}
\resizebox{0.49\hsize}{!}{\includegraphics{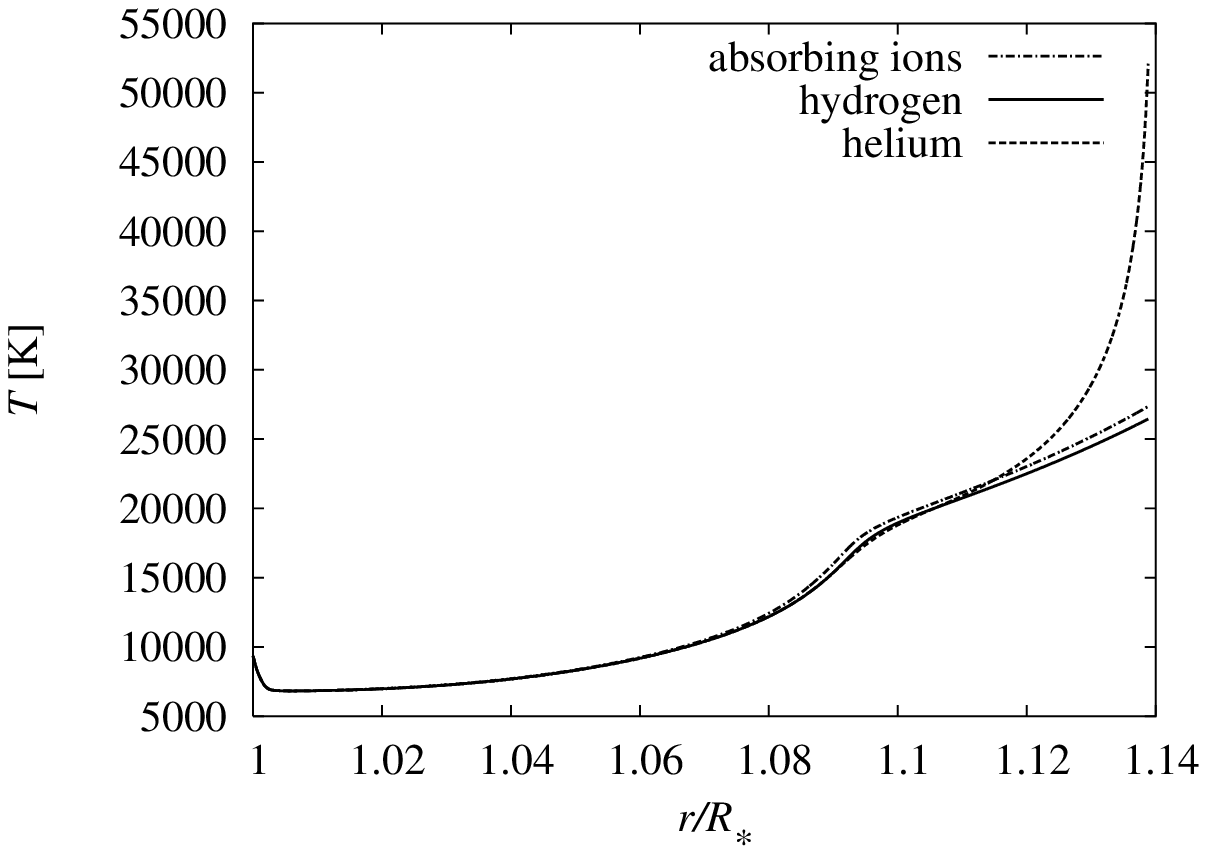}}
\resizebox{0.49\hsize}{!}{\includegraphics{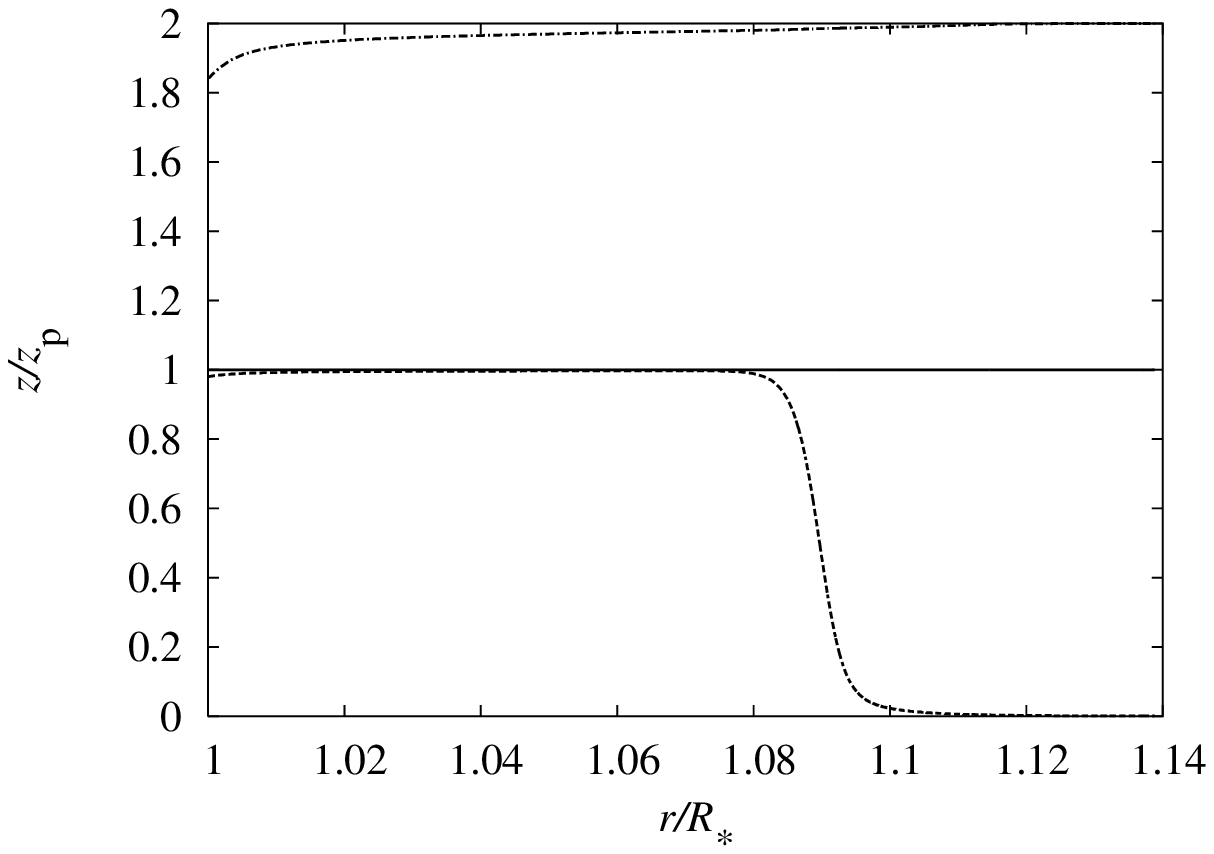}}
\caption[]{
{\bf \em Four-component wind model of a main-sequence B6 star.}
{\em Upper left panel:} Wind velocities of wind components.
Note that helium decouples from the stellar wind and subsequently
decelerates.
Helium velocity is lower that the escape velocity.
Thus, helium may fall back onto the stellar surface.
{\em Upper right panel:} Temperatures of wind components.
Stellar wind is heated mainly due to the frictional heating in the outer
parts of the wind.
{\em Lower panel:} Charges of wind components.
Note that helium recombines in the outer model region.}
\label{b6}
\end{figure}

We have selected main-sequence B6 star as a representative for 
He-poor
stars.
Again, stellar parameters are taken from Harmanec (1988) and force
multipliers are after Abbott (1982).
The plot of velocities, temperatures and charges of individual wind
components is given in Fig.\,\ref{b6}.

Apparently, in the outer model region helium atoms recombine.
Because the frictional acceleration depends on the square of helium
atomic charge (see Eqs.\,(\ref{treni},\ref{trecko})), the frictional
acceleration is not able to support helium atoms any more and helium
decouples from the flow.
The decoupling is accompanied by large frictional heating.
The helium velocity is lower than the escape velocity, thus helium may
fall back onto the stellar surface and create helium overabundance in
the stellar atmosphere.
However, this spectral type corresponds to He-poor stars, thus even in
this case the
four-component
models are not able to explain their chemical peculiarity.

\section{Conclusions and discussion}

We have calculated four-component wind models (i.e. models with
absorbing ions, hydrogen, helium, and free electrons)
applicable to
helium chemically peculiar stars.
We used our models to test
the
explanation of helium chemically peculiar stars by helium and hydrogen
decoupling in the stellar winds of these stars.
For hot B stars (in the parameter range of He-rich stars) helium is
coupled not only to metals but also to hydrogen atoms, and
it
is ionized.
Consequently, helium does not decouple from the stellar wind of He-rich
stars.
We conclude that the explanation of chemical peculiarity of He-rich
stars (HG) based on helium decoupling is
questionable.

On the other hand, for cool B stars (in the parameter range of He-poor
stars) helium may recombine and consequently fall back onto the stellar
surface.
However, this decoupling can not explain chemical peculiarity of He-poor
stars, since
it 
may produce
a surface overabundance of helium, not an underabundance, as
anticipated by HG.
 
However, there are two effects which are still unclear.
First of all, the mass-loss rates of B stars are
highly
uncertain.
Our
test calculations showed that due to lower mass-loss rate
{\em both}
hydrogen and helium may fall back onto the stellar surface in the case
of He-rich stars.
However, this effect can not help to explain the chemical peculiarity of
these stars since hydrogen and helium are well coupled even in this
case.
Another problem is the calculation of wind ionization structure.
From the theoretical point of view, as
shown by Krti\v cka \& Kub\'at (2001), the helium decoupling is possible
for artificially
lowered
helium charge.
However, our
simplified
ionization calculations (based on a nebular approximation) do not
predict such low helium charge.
We think that our calculated ionization structure is roughly correct,
but we shall perform tests
using more detailed calculations.
Both these effects, i.e. correct mass-loss rates and ionization
structure, will be incorporated in our NLTE wind code.
First results obtained using our code are currently under way
(Krti\v cka \& Kub\'at 2003).

We point out that helium diffusion
superimposed with the multicomponent
stellar wind
may explain chemical peculiarity of He-rich stars (Michaud et al. 1987).
On the other hand, mechanism of launching of He-free wind proposed by
HG may
possibly
work for some cooler stars, however probably not for He-rich stars 
(see also Kub\'at \& Krti\v cka
2004).

\begin{acknowledgements}
This work was supported by grants GA \v{C}R 205/01/0656
and 205/02/0445.
The Astronomical Institute Ond\v{r}ejov is supported by projects
K2043105 and Z1003909.
\end{acknowledgements}


\end{document}